\documentstyle[12pt]{article}
\nopagebreak

\def\3j#1#2#3#4#5#6{\mbox{$\left(\begin{array}{ccc}
#1 & #2 & #3 \\
#4 & #5 & #6
\end{array}
\right)$}}

\def\6j#1#2#3#4#5#6{\mbox{$\left\{\begin{array}{ccc}
#1 & #2 & #3 \\
#4 & #5 & #6
\end{array}
\right\}$}}

\def\eq{\begin{equation}}

\def\ee{\end{equation}}

\def\eqa{\begin{eqnarray}}

\def\eea{\end{eqnarray}}

\def\ket#1{\mbox{$\vert #1\rangle$}}

\begin{document}


\centerline{\Large{\bf Virtual Compton scattering off nuclei}}

\medskip

\centerline{\Large{\bf in the $\Delta$-resonance region}}

\vskip 1.5cm

\centerline{\large{B.~Pasquini and S.~Boffi}}

\vskip 1.0cm

\centerline{\small Dipartimento di Fisica Nucleare e Teorica, Universit\`a
di Pavia, and}

\centerline{\small Istituto Nazionale di Fisica Nucleare, 
Sezione di Pavia, Pavia, Italy}

\vskip 1.5cm


\begin{abstract}

\noindent Virtual Compton scattering in the $\Delta$-resonance region is
considered in the case of a target nucleus. The discussion involves
generalized polarizabilities and is developed for zero-spin nuclei, focusing
on the new information coming from virtual Compton scattering in comparison
with real Compton scattering.

\end{abstract}

\bigskip



\vskip 1.5cm
\clearpage

Virtual Compton scattering (VCS) as a process where a space-like virtual
photon is scattered into a real photon has recently attracted much
interest  in the case of a nucleon target~\cite{[Audit]}. In the
hard-scattering regime it allows a stringent test of perturbative
QCD~\cite{[Farrar]}, while below pion threshold it opens the possibility of
measuring new electromagnetic observables which generalize the usual
electric and magnetic polarizabilities~\cite{[Guichon]}. These observables
are clean probes of the non-perturbative structure of the nucleon and are
complementary to the elastic form factors. 

Compared to real Compton scattering, VCS experiments are much more difficult
because of the necessary interplay between the Bethe-Heitler (BH) process
and the full VCS process. The radiative corrections must be calculated with
sufficient reliability in order to be able to extract the interesting
information on the target structure. On the other hand, the increased
number of available observables suggests that VCS could be a promising
field of investigation also with nuclear targets.

In real Compton scattering on nuclei with polarized photons two structure
functions contribute to the cross section~\cite{[Vesper],[Pasquini]}. The
first structure function, $W_{\rm T}$, is the same quantity determined by
scattering of unpolarized photons and is the incoherent sum of
photon-helicity flip and non-flip contributions~\cite{[KMO]}. 
The second structure
function, $W_{\rm TT}$, contains interference contributions from helicity
flip and non-flip amplitudes. Thus their separated determination is
important for disentangling different reaction mechanisms. Such a
separation has been achieved in a recent experiment on $^4$He in the
$\Delta$-resonance region~\cite{[Schaerf]} adding further information to
the precise data recently obtained at Mainz with tagged
photons~\cite{[Selke]}. In this region $W_{\rm T}$ and $W_{\rm TT}$ 
are differently related to the resonant and background
contributions. In a recent analysis of the data within the frame of the
$\Delta$-hole model~\cite{[Pasquini]} a rather satisfactory agreement
between theory and data is obtained under resonance conditions, while below
the $\Delta$-resonance energy discrepancies are found at backward angles
which can be ascribed to some lacking background mechanism.

The independent variation of energy and momentum transfer available in
electron scattering makes VCS in the same energy region a suitable tool to
investigate the relative importance of the longitudinal/transverse
components of the current and the behaviour of background contributions as 
a function of the photon momentum transfer. In this letter we study the
general nuclear VCS response and discuss the particular case of zero-spin
nuclei, ignoring the photon bremsstrahlung contribution of the electrons in
the BH process.

Assuming the validity of the Born approximation, an incident electron with
four-momentum $k^{\mu}=(E,\vec{k})$ is scattered to a final four-momentum
$ k'^{\mu}=(E',\vec{k}')$ exchanging with the nuclear target a virtual
photon with four-momentum $q^{\mu}=k'^{\mu}-k^{\mu}=(\omega,\vec{q})$. The
elettroexcitation of the nucleus from the initial state  $\ket{J_{\rm i}
M_{\rm i}}$ is followed by emission of a photon
$q'^{\mu}=(\omega'=\omega,\vec{q'})$ with the nuclear transition to the
final state $\ket{J_{\rm f}M_{\rm f}}\equiv \ket{J_{\rm i}M_{\rm i}}$.
The full VCS cross section involves the calculation of
\eq
M=\sum_{\lambda, \lambda',\bar{\lambda}}\sum_{M_{\rm i}, M_{\rm f}}
(T^{\lambda'\lambda}_{M_{\rm f}M_{\rm i}})\,
L_{\lambda\bar{\lambda}}\,
(T^{\lambda'\bar{\lambda}}_{M_{\rm f}M_{\rm i}})^{*},
\label{eq:crosssec} 
\ee
where $L_{\lambda\bar{\lambda}}$ is the lepton tensor and 
$T^{\lambda'\lambda}_{M_{\rm f}M_{\rm i}}$
is the nuclear scattering amplitude of a virtual photon with polarization 
$\lambda=0,\pm 1$ into a real photon with polarization $\lambda '=\pm 1$.
The VCS amplitude $T^{\lambda'\lambda}_{M_{\rm f}M_{\rm i}}$
can be conveniently described in terms of nuclear polarizabilities~\cite{[ad]}
 as a direct generalization of the formalism developed in~\cite{[Pasquini]} for
real Compton scattering. Due to the nature of the virtual photon, in
addition to the contribution of the $M^{\nu=1,\vert\lambda\vert=1}=M$
(magnetic) and $M^{\nu=2,\vert\lambda\vert=1}=E$ (electric) multipoles,
the multipole expansion of the absorbed photon includes the
$M^{\nu=0,\lambda=0}=C$ (Coulomb or longitudinal) multipole.  As a
consequence, the definition of the nuclear polarizabilities of eq. $(1)$ 
of ref.~\cite{[Pasquini]} is generalized by 
\eqa
& &P_{J}(M^{\nu'\lambda'}L',M^{\nu\lambda} L;q'^{\mu},q^{\mu}) 
= \nonumber \\
& & \nonumber \\ 
& &\quad =  (-)^{L+L'-J_{\rm f}} \,{\hat L}^2\, {\hat
{L'}}^2 \sum_{M_{\rm i} M_{\rm f}}\sum_{MM'm} {\textstyle {1\over 4}}
 \sum_{\lambda=0,\pm 1}\sum_{\lambda'=\pm 1}
\lambda^\nu {\lambda'}^{\,\nu'}(1+\delta_{\lambda,0})
\nonumber \\
& & \nonumber \\
& &
\qquad \times 
(1-\delta_{\nu,0}\,\delta_{|\lambda|,1})
\,(-)^{M_{\rm i}}
\3j{I_{\rm f}}{J}{I_{\rm i}}{-M_{\rm f}}{m}{M_{\rm i}}\,
\3j{L}{L'}{J}{M}{M'}{-m} \nonumber \\
& & \nonumber \\
& &\qquad \times \frac{1}{(8\pi^2)^2}
\int{\rm d}R \int{\rm d}R'\,{\cal D}^{L'*}_{M'-\lambda'}(R')\,
T^{\lambda'\lambda}_{M_{\rm i}M_{\rm f} }(q'^{\mu},q^{\mu})\,
{\cal D}^{L*}_{M\lambda}(R).
\label{eq:polgen}\\ \nonumber
\eea
The total angular momentum $J$ transferred to the nucleus is limited
by the Clebsch-Gordan coefficients as follows
\eq  
\vert J_{\rm i}-J_{\rm f}\vert\leq J\leq J_{\rm i}+J_{\rm f},  
\qquad \vert L-L'\vert \leq J\leq L+L',
\label{eq:jsel}
\ee
while parity conservation requires 
\eq 
P_{J}(M^{\nu'\lambda'}L',M^{\nu\lambda} L;q'^{\mu},q^{\mu})=0 
\hbox{\ } \hbox{\ } \hbox{\ }
\mbox{if $(-1)^{L+\nu+L'+\nu'}\neq\pi_i\pi_f $},\label{eq:psel}
\ee
where the parities of the initial and final states of the nucleus have been 
denoted by $\pi_i$ and $\pi_f$, respectively.

In the case of unpolarized electrons, eq.~(\ref{eq:crosssec}) can be 
rewritten in terms of four response functions of the nucleus
\eq
M = L_{00}W_{\rm L} + L_{11}W_{\rm T} + L_{01}W_{\rm LT}\cos\alpha 
+ L_{1-1}W_{\rm TT}\cos 2\alpha, 
\ee
where $\alpha$ is the azimuthal angle of the emitted photon with respect to
the electron scattering plane. In contrast to the case of real polarized
photons, where only the pure transverse $W_{\rm T}$ and the 
transverse-transverse
interference $W_{\rm TT}$ responses occur, now there are also the pure
longitudinal $W_{\rm L}$ and the longitudinal-transverse interference 
$W_{\rm LT}$
structure functions.

Due to the selection rules~(\ref{eq:jsel}) and (\ref{eq:psel}), for a
zero-spin target nucleus only the scalar polarizabilities 
$P_{0}(EL,EL;q'^{\mu},q^{\mu})$,  $P_{0}(ML,ML;q'^{\mu},q^{\mu})$ and 
$P_{0}(EL,CL;q'^{\mu},q^{\mu})$ survive. Then the decomposition of 
$W_{\rm L}$ and $W_{\rm LT}$ in terms of polarizabilities is given by 

\eq
W_{\rm L} = \sum_{\lambda'=\pm 1}
\left\vert\sqrt{2} \sum_{L}\,(-)^{L}\,{\hat L}^{-1}
d^L_{0,\lambda}(\theta_{\gamma}) 
P_{0}(EL,CL;q'^{\mu},q^{\mu})\right\vert^2,
\label{eq:wll}
\ee

\eqa
W_{\rm LT}  &=& \sqrt{2}\sum_{L, \bar{L}}\,(-)^{L+\bar{L}}\,
{\hat L}^{-1}\,\hat{\bar{L}}^{-1}\nonumber \\
& &\quad\times
\sum_{\lambda,\lambda'=\pm 1} \lambda\, 
d^L_{0,\lambda'}(\theta_{\gamma}) \,
d^{\bar{L}}_{\lambda,\lambda'}(\theta_{\gamma}) 
\, 2{\rm Re}\Big\{
P^{}_{0}(EL,CL;q'^{\mu},q^{\mu})^{}\nonumber\\
& &\qquad\times
\left[P_{0}(E\bar{L},E\bar{L};q'^{\mu},q^{\mu})
+\lambda\lambda' P_{0}(M\bar{L},M\bar{L};q'^{\mu},q^{\mu})\right]^{*}
\Big\}. \label{eq:wlt}
\eea

For $W_{\rm T}$ and $W_{\rm TT}$ the same expressions
in eqs. (10) and (11) of ref.~\cite{[Pasquini]} hold, where now
${\cal D}^L_{\pm1,\pm1}(R')$ are replaced by the reduced rotation matrices 
$d^L_{\pm 1,\pm 1}(\theta_{\gamma})$. 

In principle, VCS allows us to access new information with respect to real
Compton scattering through the scalar 
polarizability $P_0(EL,CL;q'^{\mu},q^{\mu}),$ 
which involves the coupling between the longitudinal
virtual photon and the nuclear transition operator. 
This is achieved either by a Rosenbluth separation of $W_{\rm L}$ and 
$W_{\rm T}$ in 
parallel conditions ($\theta_{\gamma}=0$) or by looking at the left-right 
asymmetry ($\cos\alpha=\pm 1$) to extract $W_{\rm LT}.$   

In the
$\Delta$-resonance region we are facing a situation similar to that
occurring when studying the $E2/M1$ ratio for the $\Delta$ excitation in
pion photoproduction, where the longitudinal-transverse $W_{\rm LT}$ 
structure function emphasizes the role of the small $C2/E2$ transition
through the amplifying factor given by the interfering $M1$
transition~\cite{[Laget]}.

Within the frame of the modified version of the $\Delta$-hole
model described in ref.~\cite{[Pasquini]}, the nuclear transition operator
is modelled by one-body interaction mechanisms between the incident photon
and the single nucleon. As a consequence, the nuclear polarizabilities can
be directly derived from the nucleon polarizability operators 
$P_j^{\alpha}(M^{\nu'\lambda'}l',M^{\nu\lambda}l;
q'^{\mu},q^{\mu})$, and in the particular case of longitudinal photons one
obtains 
\eqa
& &P_{0}(EL,CL;q'^{\mu},q^{\mu}) = \nonumber\\
& &{}= \sqrt{\pi}\sum_{n n'}\sum_{l l'}
\sum_{j  m \lambda'}\sum_{\alpha=1}^{A}\, 
{\rm i}^{n + n'}\hat{n}^2\hat{n}'^{\,2}
\hat{L}^3\hat{j}(-1)^{l'+L-m}j_{n'}(q'r_{\alpha})j_{n}(qr_{\alpha})
Y_{j,-m}(\hat{r}_{\alpha})\nonumber\\
& &\nonumber\\
& &\quad
\times 
\3j{n}{n'}{j}{0}{0}{0}
\3j{n}{l}{L}{0}{0}{0}
\3j{n'}{l'}{L}{0}{-\lambda'}{\lambda'}\nonumber\\
& & \nonumber\\
& & \quad
\times 
\6j{n}{n'}{j}{l'}{l}{L}
\left[ P_{j}^{\alpha}(E{l'},C{l};q'^{\mu},q^{\mu})
+ \lambda' P_{j}^{\alpha}(M{l'},C{l};q'^{\mu},q^{\mu})\right],
\nonumber\\
& &\label{eq:polel}
\eea
where the nucleon polarizabilities $P_{j}^{\alpha}$ satisfy the
same selection rules of eqs. (\ref{eq:jsel}) and (\ref{eq:psel}) applied to
the nucleon quantum numbers.

Resonant multipoles for the $N\Delta$ transition are $M1$, $E2$, and
$C2$. The corresponding transition currents can be cast in the form
of, e.g., ref.~\cite{[Lagetnp]}. Besides the obviously dominant
pure $M1$ contribution, for zero-spin nuclei only the 
longitudinal $P_{0}^{\alpha}(E2,C2;q'^{\mu},q^{\mu})$ and
transverse $P_{0}^{\alpha}(E2,E2;q'^{\mu},q^{\mu})$
are different from zero.

Background contributions come from the Kroll-Ruderman term and the
seagull term in the two-photon amplitude. The latter is pure
transverse. From the Kroll-Ruderman term one obtains
the elementary polarizabilities $P_0^{\alpha}(E1,E1;q'^{\mu},q^{\mu})$
and $P_1^{\alpha}(E1,C1;q'^{\mu},q^{\mu})$. Specializing to zero-spin
nuclei, it turns out that only the elementary scalar polarizability
contributes. Then, for the background one has 
$P_{0}(EL,CL;q'^{\mu},q^{\mu})=0$.
As a result, in the above model only small contributions to $W_{\rm L}$
and $W_{\rm LT}$ are possible from the resonant $E2$ and $C2$ multipoles.
In a more refined model other background terms could be included giving
rise, e.g., also to possible contributions from
$P_0^{\alpha}(E1,C1;q'^{\mu},q^{\mu})$. However, this terms are expected
to be of even smaller size.

In fig. 1 the structure functions $W_{\rm T}$, $W_{\rm TT}$ and $W_{\rm LT}$ 
are given for $^4$He across the $\Delta$-resonance region. The
behaviour of $W_{\rm T}$ and $W_{\rm TT}$ is an extrapolation of what can be
calculated~\cite{[Pasquini]} and observed~\cite{[Schaerf]} in real
Compton scattering. The new structure function $W_{\rm LT}$ is rather small
and peaked in the forward hemisphere. Its separation 
requires high-precision experiments to get rid of the overwhelming
photon bremsstrahlung contribution in the observed cross section.


\clearpage


\centerline{\bf Figure captions}

\medskip

Fig. 1. Angular dependence of the structure functions $W_{\rm TT},$
$W_{\rm LT}$ e $W_{\rm T}$  
for virtual Compton scattering off $^4$He at an excitation energy of 310
MeV. Dotted, dot-dashed, dashed and solid lines for an incident photon
momentum of 330, 380, 430 and 480 MeV, respectively.

\smallskip


\begin{thebibliography}{99}

\bibitem{[Audit]}
{G.~Audit et al., CEBAF proposal PR-93-050 (1993); MAMI proposal A1/1-95
(1995).} 

\bibitem{[Farrar]}
{G.R.~Farrar and H.~Zhang, Phys. Rev. D41 (1990) 3348.}

\bibitem{[Guichon]}
{P.A.M.~Guichon, G.O.~Liu, and A.W.~Thomas, Nucl. Phys. A591 (1995) 606.}

\bibitem{[Vesper]}
{J.~Vesper, D.~Drechsel and N.~Ohtsuka, Nucl. Phys. {A466} (1987) 652.}

\bibitem{[Pasquini]}
{B.~Pasquini and S.~Boffi, Nucl. Phys. {A598} (1996) 485.}

\bibitem{[KMO]}
{J.H.~Koch, E.J.~Moniz and N.~Ohtsuka, Ann. Phys. (NY) {154} (1984) 99.}

\bibitem{[Schaerf]}
{C.~Schaerf, in Conference on Perspectives in Nuclear Energy at
Intermediate Energies, ed. S.~Boffi, C.~Ciofi degli Atti and M.M.~Giannini
(World Scientific, Singapore, 1996) p. 199.}

\bibitem{[Selke]}
{O.~Selke et al., Phys. Lett. {B369} (1996) 207.}

\bibitem{[ad]}
{H.~Arenh\"ovel and D.~Drechsel, Nucl. Phys. {A233} (1974) 153.}

\bibitem{[Laget]}
{J.-M. Laget, in New Vistas in Electro-Nuclear Physics, ed. by
E.L.~Tomusiak, H.S.~Caplan and E.T.~Dressler (Plenum Publ. Co., New York,
1986) p. 361; Can. J. Phys. {62} (1984) 1046.}

\bibitem{[Lagetnp]}
{J.-M. Laget, Nucl. Phys. {A481} (1988) 765.}

\end{thebibliography}
\end{document}